\documentstyle[12pt,epsfig]{article}
\advance\textheight by \topskip
\begin{document}
\tolerance=100000
\thispagestyle{empty}
\setcounter{page}{0}

\newcommand{\be}{\begin{equation}}
\newcommand{\ee}{\end{equation}}
\newcommand{\br}{\begin{eqnarray}}
\newcommand{\er}{\end{eqnarray}}
\newcommand{\ba}{\begin{array}}
\newcommand{\ea}{\end{array}}
\newcommand{\bi}{\begin{itemize}}
\newcommand{\ei}{\end{itemize}}
\newcommand{\bn}{\begin{enumerate}}
\newcommand{\en}{\end{enumerate}}
\newcommand{\bc}{\begin{center}}
\newcommand{\ec}{\end{center}}
\newcommand{\ul}{\underline}
\newcommand{\ol}{\overline}
\newcommand{\ar}{\rightarrow}
\newcommand{\sm}{${\cal {SM}}$}
\newcommand{\susy}{{{SUSY}}}
\newcommand{\Dir}{\kern -6.4pt\Big{/}}
\newcommand{\Dirin}{\kern -10.4pt\Big{/}\kern 4.4pt}
\newcommand{\DDir}{\kern -10.6pt\Big{/}}
\newcommand{\DGir}{\kern -6.0pt\Big{/}}
\def\gluino{\ifmmode{\mathaccent"7E g}\else{$\mathaccent"7E g$}\fi}
\def\photino{\ifmmode{\mathaccent"7E \gamma}\else{$\mathaccent"7E \gamma$}\fi}
\def\mgluino{\ifmmode{m_{\mathaccent"7E g}}
             \else{$m_{\mathaccent"7E g}$}\fi}
\def\taugluino{\ifmmode{\tau_{\mathaccent"7E g}}
             \else{$\tau_{\mathaccent"7E g}$}\fi}
\def\mphotino{\ifmmode{m_{\mathaccent"7E \gamma}}
             \else{$m_{\mathaccent"7E \gamma}$}\fi}
\def\ML{\ifmmode{{\mathaccent"7E M}_L}
             \else{${\mathaccent"7E M}_L$}\fi}
\def\MR{\ifmmode{{\mathaccent"7E M}_R}
             \else{${\mathaccent"7E M}_R$}\fi}

\def\Ord{\buildrel{\scriptscriptstyle <}\over{\scriptscriptstyle\sim}}
\def\OOrd{\buildrel{\scriptscriptstyle >}\over{\scriptscriptstyle\sim}}
\def\pl #1 #2 #3 {{\it Phys.~Lett.} {\bf#1} (#2) #3}
\def\np #1 #2 #3 {{\it Nucl.~Phys.} {\bf#1} (#2) #3}
\def\zp #1 #2 #3 {{\it Z.~Phys.} {\bf#1} (#2) #3}
\def\pr #1 #2 #3 {{\it Phys.~Rev.} {\bf#1} (#2) #3}
\def\prep #1 #2 #3 {{\it Phys.~Rep.} {\bf#1} (#2) #3}
\def\prl #1 #2 #3 {{\it Phys.~Rev.~Lett.} {\bf#1} (#2) #3}
\def\mpl #1 #2 #3 {{\it Mod.~Phys.~Lett.} {\bf#1} (#2) #3}
\def\rmp #1 #2 #3 {{\it Rev. Mod. Phys.} {\bf#1} (#2) #3}
\def\sjnp #1 #2 #3 {{\it Sov. J. Nucl. Phys.} {\bf#1} (#2) #3}
\def\cpc #1 #2 #3 {{\it Comp. Phys. Comm.} {\bf#1} (#2) #3}
\def\xx #1 #2 #3 {{\bf#1}, (#2) #3}
\def\preprint{{\it preprint}}

\begin{flushright}
{Cavendish-HEP-96/8}\\
{DFTT 46/96}\\ 
{UGR-FT-66}\\
{MZ-THEP-96-29}\\ 
{June 1996\hspace*{.5 truecm}}\\
%{Modified August 1996}\\
\end{flushright}

\vspace*{\fill}

\begin{center}
{\Large \bf 
Heavy flavour tagging and QCD tests\\
in four-jet events at LEP1}\\[0.5 cm]
{\large S. Moretti$^{a,b}$, R. Mu\~noz-Tapia$^c$ and 
           J.B. Tausk$^{d}$}\\[0.4 cm]
{\it a) Cavendish Laboratory, University of Cambridge,}\\
{\it Madingley Road, Cambridge CB3 0HE, UK.}\\[0.25cm]
{\it b) Dipartimento di Fisica Teorica, Universit\`a di Torino,}\\
{\it and I.N.F.N., Sezione di Torino,}\\
{\it Via Pietro Giuria 1, 10125 Torino, Italy.}\\[0.25cm]
{\it c) Dpto F\'{\i}sica Te\'orica y del Cosmos,}\\
{\it Universidad de Granada, Granada 18071, Spain.}\\[0.25cm]
{\it d) Johannes Gutenberg-Universit\"at, Institut f\"ur Physik (THEP),}\\
{\it    Staudingerweg 7, D-55099 Mainz, Germany.}\\[0.5cm]
\end{center}
\centerline{\sl Contributed to the 1996 International Conference on High 
           Energy Physics,}
\centerline{\sl 25-31 July 1996, Warsaw, Poland (paper pa04--28).}
\vskip0.5cm
\noindent
%\centerline{\it Proposed speaker: S. Moretti.}
\vspace*{\fill}

\begin{abstract}
{\noindent 
\small 
We present a theoretical review of the possibilities offered by
heavy flavour tagging in testing QCD features in 4-jet events at LEP1.
Attention is focused mainly on the properties of the angular
variables which are used in the experimental analyses to measure
the colour factors of the strong interactions.
We also comment on a possible settlement of
the ongoing controversy about the existence in LEP1 data
of SUSY events involving light gluinos.
Integrated and differential rates of interest to phenomenological
analyses are given and various tagging procedures are discussed.
}
\end{abstract}
\vskip1.0cm
\hrule
\vskip0.25cm
\noindent
E-mails: moretti@hep.phy.cam.ac.uk,
         rmt@ugr.es,
         tausk@vipmzw.physik.uni-mainz.de.

\vspace*{\fill}
\newpage

\subsection*{1. Introduction}

The LEP1 era has now ended. Regarding QCD tests, 
the legacy passed on by the 
experiments ALEPH, DELPHI, L3 and OPAL is indeed a convincing indication
that the $SU(3)$ gauge group of the Standard Model (\sm) is the theory that 
describes the strong interactions \cite{Hebbeker}.
By studying the process 
$e^+e^-\ar Z \ar\mathrm{hadrons}$ decisive results have been achieved:
the QCD coupling `constant' $\alpha_s$ has been determined from jet
rates and shape variables; both its 
flavour independence and running 
have been verified; multi-jet 
distributions reproduce QCD predictions to second order in $\alpha_s$;
many models proposed as alternatives to ordinary QCD
have been ruled out; the triple gluon vertex coupling has been
verified to be in agreement with QCD;
finally, the colour factors that determine the gauge
group responsible for strong interactions have been measured 
\cite{colfac}. However, as far it concerns
the latter measurements, somewhat less conclusive results have been obtained.

In the `colour factor analyses' in 4-jet samples,  
the basic idea 
is to measure the fundamental colour factors of QCD, 
that is, $C_A$ and $C_F$ and $T_F$, which are determined
from the $SU(N_C)$ generators $(T^a)_{ij}$ and from the structure constants
$f^{abc}$. Explicitly, one obtains 
$C_A=N_C$,
$C_F=(N_C^2-1)/2N_C$,
and $T_F=1/2$.
In ordinary QCD (i.e., $N_C=3$): $C_A=3$ and $C_F=4/3$.
In order to determine $C_A$, $C_F$ and $T_F$\footnote{Better, ratios among 
them: e.g., $C_A/C_F$ and $T_R/C_F$, where $T_R=N_FT_F$, 
with $N_F$ number of active flavours.},  
one compares the theoretical predictions to the data by
leaving the colour factors as free parameters to be determined by
a fit. The comparison is made by resorting to differential spectra of
variables based on angular correlations between jets, which are usually 
ordered in energy and where the two most energetic jets
are `identified' with the primary quarks (i.e., from the $Z$ decay). 
Some of the most widely used 
are the Bengtsson-Zerwas angle $\chi_{BZ}$, the
K\"orner-Schierholz-Willrodt angle $\Phi_{KSW}$,
the `modified' Nachtmann-Reiter angle $\theta_{NR}^*$ \cite{angles}
and the angle
between the two least energetic jets $\theta_{34}$ (i.e.,
from the gluon splitting).
The distributions in
these quantities are different for $Q\bar{Q}gg$ and
$Q\bar{Q}q\bar{q}$ events. This is decisive as, in general, 
the main difference between QCD and alternative models is the predicted
relative contributions of the two 
subprocesses\footnote{E.g., the $2Q2q$ contribution to the total 4-jet rate
at the $Z$ peak is only about $5\%$ in QCD but up to 30\% 
in certain Abelian models \cite{nonAbel}.}. This
leads to different shapes of the angular distributions of the full
4-jet sample and thus
to different predictions for the values
of $C_A$, $C_F$ and/or $T_F$ (see, e.g., Ref.~\cite{nonAbel}).
Ultimately, in fact, $C_F$, $C_A$ and
$T_F$ represent nothing else than
the relative strengths of the interactions $q\ar qg$, $g\ar gg$
and $g\ar q\bar q$, respectively.

The present status of the colour factor analyses \cite{colfac,OPAL} is that,
although the experimental measurements agree well with ordinary 
QCD, it is not possible to rule out its Supersymmetric
version. In fact, such a theory includes in the spectrum of the
fundamental particles the spin 1/2 partner
of the gluon, the {\sl gluino} \gluino, which could be produced at
LEP1 in $Q\bar Q
\gluino\gluino$ events (via $g^*\ar \gluino\gluino$),
provided it is not too heavy. Since
gluinos are coloured fermions, their contribution would enhance the part of the
4-jet cross section with angular dependence similar to that of
$2Q2q$ events.
Naively, one could alternatively say 
that $N_F$ is (apparently)
increased, such that
an experimental measurement could reveal
\susy\ signals in the form of an enhancement of
$T_R$.
In detail, gluinos with a mass $\OOrd2$ GeV yield an expectation 
value for $T_F/C_F$
within one standard deviation of the measured one \cite{OPAL}. 

Like the QCD gauge boson, the 
gluino is neutral, it is its own anti-particle (i.e.,
it is a Majorana fermion) and its coupling to ordinary matter is precisely
determined in terms of the QCD colour matrices and 
$\alpha_s$ \cite{Nilles}.
However, whereas $m_g\equiv0$,
the gluino mass \mgluino\ is an arbitrary
parameter, and so is the lifetime \taugluino.
In particular, 
in the theory it is natural for gluinos to be much lighter (a few GeV)
than squarks
if their mass is induced radiatively and 
if dimension-3 \susy\ breaking operators are
absent from the low energy dynamics \cite{Masiero-Farrar-Quarks94}.
Thus, if gluinos are so light, they should be produced at LEP1 \cite{epem}.
Recently, there has been a renewed interest in this possibility, especially
motivated by the (small) discrepancy between the value of $\alpha_s$ 
determined by
low energy deep-inelastic lepton-nucleon scattering and the one measured
by the $e^+e^-$ CERN experiments \cite{alphas}.
In this respect, 
it has been speculated that the evolution of $\alpha_s$
is being slowed down by a contribution to the $\beta$ function
due to light gluinos.
The latest experimental constraints still allow for the existence of relatively
long lived and light gluinos, in the parameter 
regions \cite{Farrar-Kileng-Osland}: 
(i) $\mgluino\Ord1.5$ GeV and $\taugluino\Ord10^{-8}$ s; 
(ii) $\mgluino\OOrd4$ GeV and $\taugluino\OOrd10^{-10}$ s. 

In our opinion, if one adopts energy ordering of the jets (as it has been 
done so far), 
the effectiveness of the experimental analyses in putting
stringent bounds on the measured values of $C_A$, $C_F$ and $T_R$ 
is largely reduced. In fact,
the angular variables introduced above are most
useful for emphasizing QCD features if one can distinguish
between quark and gluon jets and assign the momenta of the final
states to the corresponding particles\footnote{When ordering in 
energy, one has to bear in mind that in $Z\ar Q\bar Qgg$ decays 
for only half of the events the two
lowest energy partons are both gluons \cite{nonAbel} !}.
It was this consideration that motivated two of us 
to advocate in Ref.~\cite{ioebas}
the use of 4-jet samples with two of the jets
clearly identified as originating from heavy flavour quarks. 
In this way, it is at least possible to distinguish some
quark jets, namely those which originate from $b$-quarks, from gluon
jets, thus identifying
the particles in the final states and defining the angular variables in the 
proper manner. A direct consequence of this approach is
that one gains a greater discriminating power between events with
different angular behaviour. Furthermore, an intrinsic advantage of the
heavy flavour selection is that the
$Q\bar Qgg$ event rates are reduced 
more (by a factor
of 5) than the $Q\bar Q q\bar q$ ones (which are divided
by 3), such that their relative differences 
can be more easily studied.
The final aim is to assess whether with 
$b$-tagging techniques one is able to reduce the
confidence level regions around the experimental measurements, in such
a way to eventually shed new light on the possibility that \susy\ signals
are present in LEP1 data.

Such techniques have been rapidly developed by the 
LEP1 collaborations in the past years \cite{btagg},
and have become an excellent instrument to study 
heavy flavour physics. Their main features are well summarized, e.g., 
in Refs.~\cite{Squarcia}: in particular, we remind the reader 
that the method offering 
the best performances is the lifetime tagging by 
detecting a secondary decay $\mu$-vertex. 
Yet, the crucial point in $b$-tagging-based analyses 
is that the total statistics
in $2b2\mbox{jet}$ events is greatly reduced compared to that available
when no quark identification is exploited, not merely because
only flavour combinations involving $b$'s are retained
(and these are even mass suppressed \cite{MEs}) but
also because the (single) tagging efficiency at LEP1 is rather
low, $\varepsilon\Ord40\%$ (for purities close to 1). 
Typically, one has to consider that
the initial rate of $2b2\mbox{jet}$ events from pure QCD is of the order
${\cal O}(8000)$ events per experiment: for
$\approx 4\times10^{6}$ hadronic $Z$ decays, 
$\Gamma_{b\bar b}/\Gamma_{\mathrm{had}}\approx22\%$, $R_4\approx10\%$ and
$\varepsilon\approx30\%$, where $R_4$ is the 4-jet fraction, and
considering two tags. Note 
that the statistics could well be increased
by considering the data of all experiments altogether and/or by relaxing
the requirements of high $b$-purity (see discussion in Ref.~\cite{ioebas}).
Indeed, experimental studies of $b$-tagged 4-jet events have been
already reported \cite{done} (showing that a considerable reduction of
the systematics can be achieved) and others are 
in progress \cite{progress}.

Independently of the results of the new analyses,
there are however other
possibilities offered by the $\mu$-vertex devices to settle
down the debate about colour
factor measurements at LEP1. This is
clear if one considers that long-lived gluinos might yield 4-jet events 
with detectable displaced 
vertices, as already recognised in Ref.~\cite{vertex} and 
exploited in Ref.~\cite{Cuypers}. In the rest of 
this contribution, we briefly summarise the computational techniques,
the tagging procedures and the main results already reported in 
Refs.~\cite{ioebas,gluino}. 
 
\subsection*{2. Calculation} 

In carrying out the analyses described here we made use of the 
{\tt FORTRAN} matrix elements 
already discussed in Ref.~\cite{MEs} and presently 
used for experimental simulations \cite{OPAL},  
upgraded with the inclusion of the gluino 
production and decay mechanisms (see also Ref.~\cite{masses}).
The programs do not contain any approximations, the intermediate
states $\gamma^*$ and $Z$ being both inserted, and the 
masses and
polarisations of all particles in the final states
(of the two-to-four body processes) retained.
The availability of the last two options is especially important
if one considers, on the one hand, 
that in $b$-tagged samples all final states are massive,
and, on the other hand, that in proceeding to experimental fits 
one could well select restricted
regions of the differential spectra of the angular variables,
where the rates are likely to strongly depend on the spin state of the 
partons\footnote{For this reason 
we have not used the results published in literature for the gluino
decay rates, as these are averaged
over the helicities of the unstable particle.
Instead we have recomputed the relevant Feynman decay amplitudes
by preserving the gluino polarisation and by matching the latter with
the corresponding one in the production process \cite{gluino}.}.
                               
In the analyses we are reporting about we have adopted four different
jet-finding procedures: the JADE scheme (J) 
and its ``E'' variation (E), 
and the most recent Durham (D) and Geneva (G) 
algorithms \cite{schemes}.
We will present rates only for some of these, as  
none of the results 
drastically depends on the choice of the jet recombination scheme and/or
the value of the jet resolution parameter, $y_{\mathrm{cut}}$.
As it concerns the exact definitions of the angular 
variables used here, these can be found in 
Ref.~\cite{ioebas}\footnote{We stress that 
the $\Phi_{KSW}^*$ angle
is actually a modification of the original one $\Phi_{KSW}$ , proposed
by two of us (see
Ref.~\cite{ioebas} for further details).}. Anyhow,
to make clear the rest of the paper, we point out the following
(with reference to the formulae of Ref.~\cite{ioebas}).
When heavy flavour identification is implied, 
labels 1 \& 2 refer to the two tagged 
jets whereas 3 \& 4 to the two remaining ones.
If no vertex tagging is assumed,
jets are labelled according to their energy, 
${E_1\ge E_2\ge E_3\ge E_4}$.
The numerical values adopted for quark masses and \sm\ parameters
can be found in Ref.~\cite{MEs}. 

\subsection*{3. Results}

\subsection*{3.1 Angular variables}

The first two figures illustrate the improvement that can be achieved
by using heavy flavour tagging (two lower plots), 
compared to the results obtained when energy ordering is used 
(two upper plots). 
Distributions are normalised to one. The label `Abelian' for the
$2Q2g$ contribution refers to rates obtained in a
particular Abelian model of QCD\footnote{We are aware
that this model has been already ruled out in other contexts \cite{ruledout},
but we regard it as a useful tool to demonstrate the
sensitivity of the introduced angular variables to the 
features of QCD.}, in which the self-interaction
of gluons is absent \cite{qed} (see Ref.~\cite{ioebas} for details). 
Note that the $Q\bar Qq\bar q$ subprocess 
is identical in both models.
The following aspects must be emphasised in Figs.~1--2.

\noindent
(i) Concerning the angles $\chi_{BZ}$ and $\Phi_{KSW}^*$, the discrimination
power between the behaviour of the $Q\bar Qgg$ and $Q\bar Qq\bar q$ 
 components slightly improves (more 
for the second variable). In the case of the first one, the original
symmetric behaviour implied by the definition is restored. 

\noindent
(ii) Another remarkable feature of Fig.~1 is that when $b$-tagging
is exploited, the differences between the Abelian and the non-Abelian
behaviour of the $2Q2g$ component are dramatically
emphasised, for both angles.
In the case of the $\chi_{BZ}$ variable these differences 
disappear when energy ordering is used. It should also be stressed that
this is not the case for the $\Phi_{KSW}^*$ angle only because we are
using here an `improved' definition of this variable (which shows
sensitivity to the gauge structure even without any flavour tagging),
the original one being less sensitive to such differences.

\noindent
(iii) Regarding the $\theta_{NR}^*$ angle, once again the original symmetry
is restored, such that this clearly helps to differentiate the 
$2Q2g$ and $2Q2q$ subprocesses. In contrast, the behaviour of the two different
gauge interactions in the $2Q2g$ subprocess remains identical.

\noindent
(iv) Finally, for the angle $\theta_{34}$,
the improvement concerns the enhanced separation between 
the two different gluon behaviours, more than a discrimination between
gluon and quark properties.

\subsection*{3.2 Mass effects}

Fig.~3 shows the effect on the relevant differential spectra 
of the inclusion of the $b$-quark mass in the computations.
The angular distributions corresponding to $b$-contributions 
are compared to those involving massless $d$-quarks
($q$ represents here a massless flavour). (Differences are even more
dramatic when comparing $b\bar b b\bar b$ and  
$d\bar d d\bar d$ distributions, though the former flavour combination
corresponds to a suppressed 
contribution.) 
In particular, it is worth noticing that the (QCD) $Q\bar Qgg$ spectra
is less affected
by mass effects than the $Q\bar Qq\bar q$ ones and that the 
Bengtsson-Zerwas and the (modified)
K\"orner-Schierholz-Willrodt angles are more mass sensitive than
the (modified) Nachtmann-Reiter and the angle
between the two untagged jets.
In general, it must be stressed that mass effects are not negligible
and should be included in phenomenological analyses exploiting
heavy flavour tagging.
In contrast, when energy ordering is adopted (that is, when all 
quark flavours are retained in the four-jet sample), the above effects are
largely reduced (see Ref.~\cite{MEs}). In this case, in fact, the most
part of quark combinations in the final state involves massless particles.
Results have been shown for the case of the E scheme, as 
this is the algorithm
for which mass suppression is somehow larger. However, similar effects are 
visible also in the other three cases: J, D and G. 

\subsection*{3.3 Gluino tagging}

When dealing with tagging a secondary vertex possibly due to 
gluino decays, several points must be addressed \cite{gluino}.
First of all, one has to confine oneself to secondary vertex 
analyses only\footnote{Thus
neglecting other forms of heavy flavour tagging: such as
the high
$p_T$ lepton method, as gluinos do not decay semileptonically.}, however 
this technique 
has a larger efficiency than any other method
\cite{Squarcia}.
Second, the vertex has to be inside the detectors, so that only gluinos
with $\taugluino\Ord10^{-9}~\mathrm{s}$ can be
searched for \cite{Cuypers}. Nonetheless,
this represents an appealing opportunity,
as a substantial part of the mentioned $(\mgluino,\taugluino)$ window
could be covered by the experiment. 
In this respect, we exploit a sort of `degeneracy' in
lifetime between $b$-quarks and gluinos, assuming that when 
making secondary vertex tagging
one naturally includes in the $2b2\mathrm{jet}$ sample also
SUSY events, in which a \gluino\ behaves as a $b$.
We call such approach `minimal trigger' procedure, as we propose 
a tagging strategy that {does not take} into account
any of the possible differences between gluinos and $b$-quarks
in 4-jet events with two secondary vertices 
(thus, in the following, we will generally speak of `vertex tagging').
There are in fact at least three obvious dissimilarities.

\noindent
(i) Their charge is different, such that one could ask that the jet
showing a displaced vertex has a null charge. This would allow one
to isolate a sample of pure SUSY events. However, we remind
the reader that to measure the charge of a low energy jet in 4-jet events
has not been attempted before and that the procedure would certainly 
have very low efficiency (in isolating a very broad hadronic system in
an environment with high hadronic multiplicity).

\noindent
(ii) Gluino lifetimes much longer than that of the $b$ are still consistent
with experiment  (note that $\tau_b\approx
10^{-12}~\mbox{s}$), such that recognising a 
displaced vertex with decay length $d\gg3$ mm (that of the $b$) would
allow one to immediately identify gluinos. Unfortunately, most
of the $b$-tagged hadronic sample at LEP1 has been collected
via a bi-dimensional 
tagging (see, e.g., Ref.~\cite{bidimensional}). Thus,
different $d$'s could well appear the same on the event plane. Furthermore, 
tagging a $d> 3$ mm vertex would allow one
to separate gluinos with $\taugluino>\tau_b$, but this 
would not be helpful if $\taugluino\le\tau_b$.

\noindent
(iii) Other than in lifetime, $b$-quarks and gluinos can differ in mass
as well, such that one might attempt to exploit mass constraints to separate
SUSY and pure QCD events. However, on the one hand,
one could face a region of
$m_b$-\mgluino\ degeneracy 
and, on the other hand, one should cope with the ambiguities
related to the concept of mass as defined at parton level and
as measured at hadron level.

\noindent
We emphasise that
measuring the charge of the vertex tagged jet, attempting
to disentangle different decay lengths or measuring
partonic masses could well be
further refinements of the procedure we are
proposing, and that their exploitation certainly
does not spoil the validity of the latter.
In addition, all these aspects necessarily need a proper
experimental analysis, which is beyond the scope of a theoretical study.

The steps of our analysis are very simple. Under the assumption that
$b$'s and $\gluino$'s are not distinguishable by vertex tagging, one
naturally retains in the 
sample all SUSY events (whereas the ordinary QCD
components are reduced as mentioned earlier). 
Then, it is trivial to notice that there exist
clear kinematic differences between the $Q\bar Qgg$, $Q\bar Qq\bar q$
and $Q\bar Q\gluino\gluino$ components. This is shown in Fig.~4, where
we plot the quantities $Y_{ij}=M_{ij}^2/s\equiv (p_i+p_j)^2/s$, where $ij=12$ 
or $34$ and $s=E_{\mathrm{cm}}^2\equiv(M_Z)^2$. The behaviour of the curves
is dictated by the fact 
that gluinos are always secondary products in 4-jet events, 
whereas quarks and 
gluons are not (lower plots). 
When no vertex tagging is exploited,
such differences are washed away (upper plots).
The value $m_\gluino=5$ GeV is assumed for reference, the
shape of the distributions being qualitatively the same regardless of it. 
Therefore, if one simply asks to reject events for which, e.g.,  
$Y_{12}>0.2$  and/or $Y_{34}<0.1$, one gets for the total rates of the
three components the pattern exemplified in Fig.~5. From the drastic
predominance of $2Q2g$ events in the complete `unflavoured' sample 
(no vertex tagging, top left), one first obtains that
the total rates of ordinary QCD events are significantly reduced compared
to those of SUSY events
(after vertex tagging, top right),   
and eventually that the $Q\bar Q\gluino\gluino$ fraction
is always comparable to that of $Q\bar Qgg+Q\bar Qq\bar q$ events    
(when also the kinematic cuts are implemented, bottom                
left): and this is true {\sl independently} of the gluino mass,      
of the jet algorithm and                                                 
of the $y_{\mathrm{cut}}$ value used in the analysis                 
(see Ref.~\cite{gluino}).                                            

Therefore, after our event selection, SUSY signals would certainly
be identifiable, as a clear excess in the total number of 4-jet events
with two displaced vertices. Thus,
the final goal will be to definitely assess 
the presence of gluinos at LEP1 
or contradict the latter, at least over appropriate
regions of masses and
lifetimes\footnote{And this should certainly be done
after
the appropriate MC simulations, including the details of the detectors
and of the tagging procedure
as well as a generator where \mgluino\ and \taugluino\ enter as
free parameters to be determined by a fit.}. We finally stress
that, as we are concerned here with total rates and not with differential 
distributions, the event number should be sufficient to render the analysis
statistically significant\footnote{In this respect we acknowledge that many
of the aspects of our approach were already employed in
Ref.~\cite{Cuypers}, the tagging procedure sketched there 
being however 
well beyond the statistical possibilities of the LEP1 experiments.}
and that the usual ambiguities related to the fact that gluino effects
on the total number of 4-jet events are comparable in percentage to the
systematic uncertainties due to the jet hadronisation process and/or
the $y_{\mathrm{cut}}$ selection procedure are much less severe
in our approach\footnote{These are in fact the underlying difficulties 
of any analysis based on the `unflavoured' hadronic sample and/or the
jet energy ordering, which have not been overcome even in recent  
improved studies \cite{Murayama}.}. 
However, since the key point of the present approach is to exploit 
the $b/\gluino$ vertex degeneracy, a highly enriched heavy flavour sample 
should be selected in this case.

\subsection*{3.4 Gluino decays}

Before closing, we should mention that a further aspect must be
kept into account when attempting our analysis. 
It concerns the kinematics of
the gluino decays. In the most sponsored SUSY
framework \cite{Nilles}, the dominant gluino decay modes
are $\gluino\ar q\bar q\photino$ and $\gluino\ar g\photino$, where
$\photino$ represents a `photino' (better, 
the Lightest Supersymmetric Particle,
which is a superposition of the \susy\ partners
of the neutral gauge bosons of the theory).
Furthermore, the $q\bar q \photino$ channel
is, in general, largely dominant over the $g\photino$ mode \cite{Nilles}.

The crucial point is that in both cases the gluino decays into
a jet with missing energy. It is not our intention
to discuss the possibility
of selecting such a signature, as we are mainly concerned here with the
fact that the energy left to the hadronic system $E_h$ is above
the experimental cuts in minimal hadronic energy, which are used  
to reduce the
backgrounds (e.g.,
in Ref.~\cite{OPAL} the threshold was set equal to 3 GeV).
In Fig.~6 (first three plots) we show 
the $E_h$ spectra after the gluino decay, in both 
the channels. Two kinematic decay configurations are considered:
a massless photino, and a massive one (i.e., $\mphotino =1/2\mgluino$).
The message is that in the most likely SUSY scenario 
(i.e., three-body decay dominant and massless photino) all
gluino events should be retained in the event selection. Conversely, 
Fig.~6 illustrates the percentage of these which will pass the adopted
trigger requirements.
Finally, in the bottom right plot of Fig.~6 we show 
the dependence of the SUSY rates on the value of $\mgluino$.
Below $\mgluino\approx5$ GeV, the mass suppression
is always less than a factor of 2.

\subsection*{4. Summary and conclusions}

In this paper we have stressed the importance of using at LEP1
samples of 4-jet events,
in which two of the jets show a displaced vertex. This could help to improve 
some of the experimental tests of QCD and to 
possibly settle down the ongoing dispute about the existence 
of SUSY events in the data. 
Those presented are theoretical results, 
which should be in the end verified by detailed MC simulations. However, it
is our opinion that
they indicate that the matter raised and procedures 
similar to the ones outlined here would deserve experimental attention. 
%\subsection*{Acknowledgements}
\vskip0.15cm
\noindent
We are grateful to Ben Bullock for reading the manuscript.
We also acknowledge the use
of a {\tt FORTRAN} code produced by Kosuke Odagiri to 
evaluate the gluino decay rates. 
This work is supported 
by the MURST, the UK PPARC, 
the Spanish CICYT project AEN 94-0936,
the EC Programme HCM, 
Network ``Physics at High Energy
Colliders'', contracts CHRX-CT93-0357 DG 12 COMA (SM) and ERBCHBICHT (RMT),
and by the Graduiertenkolleg ``Teilchenphysik'' (JBT).
%\goodbreak

\vfill
\clearpage

%%%%%%%%%%%%%%%%%%%%%%%%%%%%%%%%%%%%%%%%%%%%%%%%%%%%%%%%%%%%%%%%%%%%%%%%%%
%\end{document}
%%%%%%%%%%%%%%%%%%%%%%%%%%%%%%%%%%%%%%%%%%%%%%%%%%%%%%%%%%%%%%%%%%%%%%%%%%

\begin{figure}[p]
~\epsfig{file=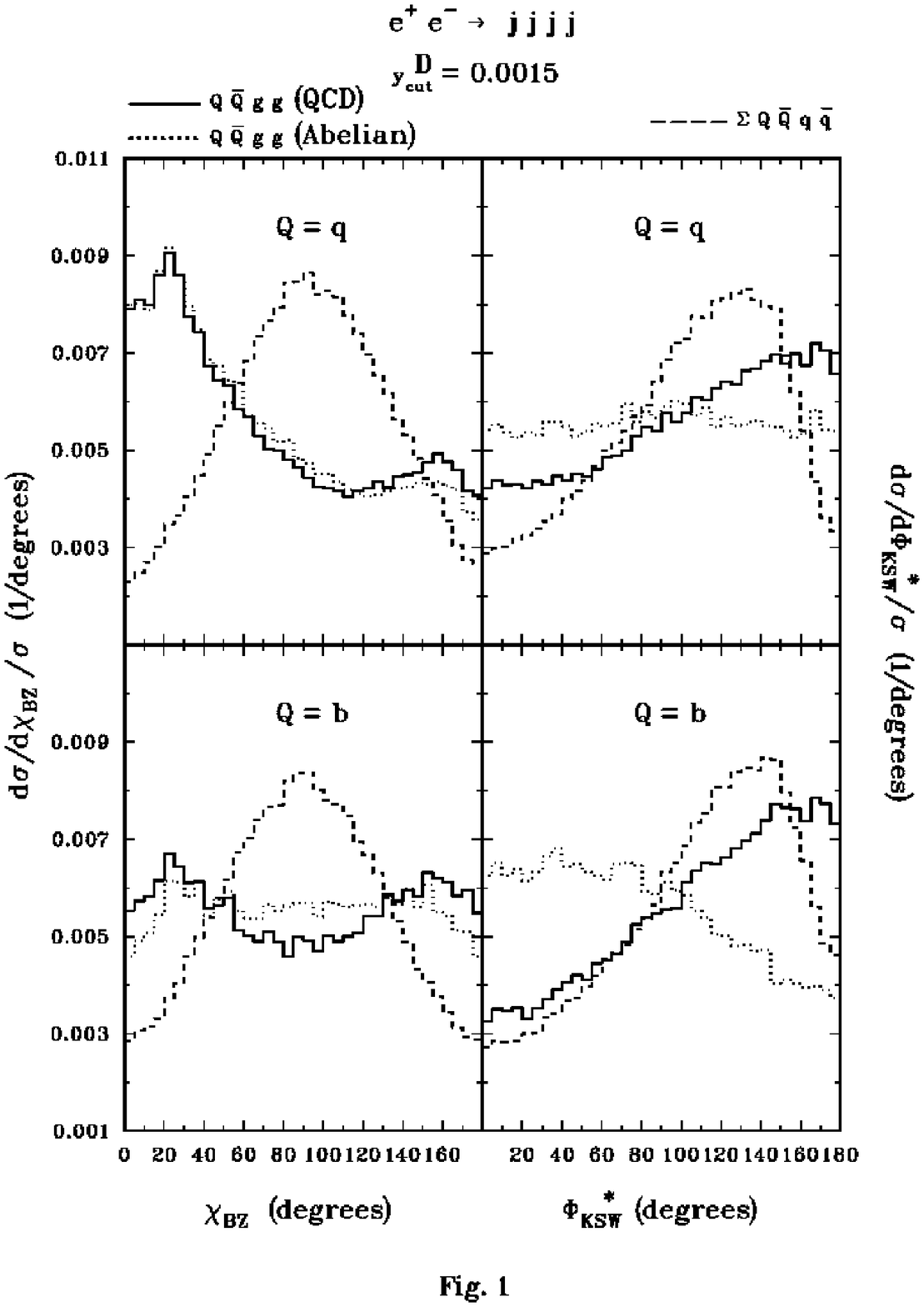,height=22cm}
\vskip0.0005cm
\noindent
{\small Distributions in the angles $\chi_{BZ}$ and $\Phi_{KSW}^*$
for the $Q\bar Qgg$ (QCD and Abelian) and $Q\bar Qq\bar q$
components in 4-jet events, in the D scheme with 
$y^D_{\mathrm{cut}}=0.0015$,
without ($Q=q$)  and with ($Q=b$) $b$-tagging.}
\end{figure}
\stepcounter{figure}
\vfill
\clearpage

\begin{figure}[p]
~\epsfig{file=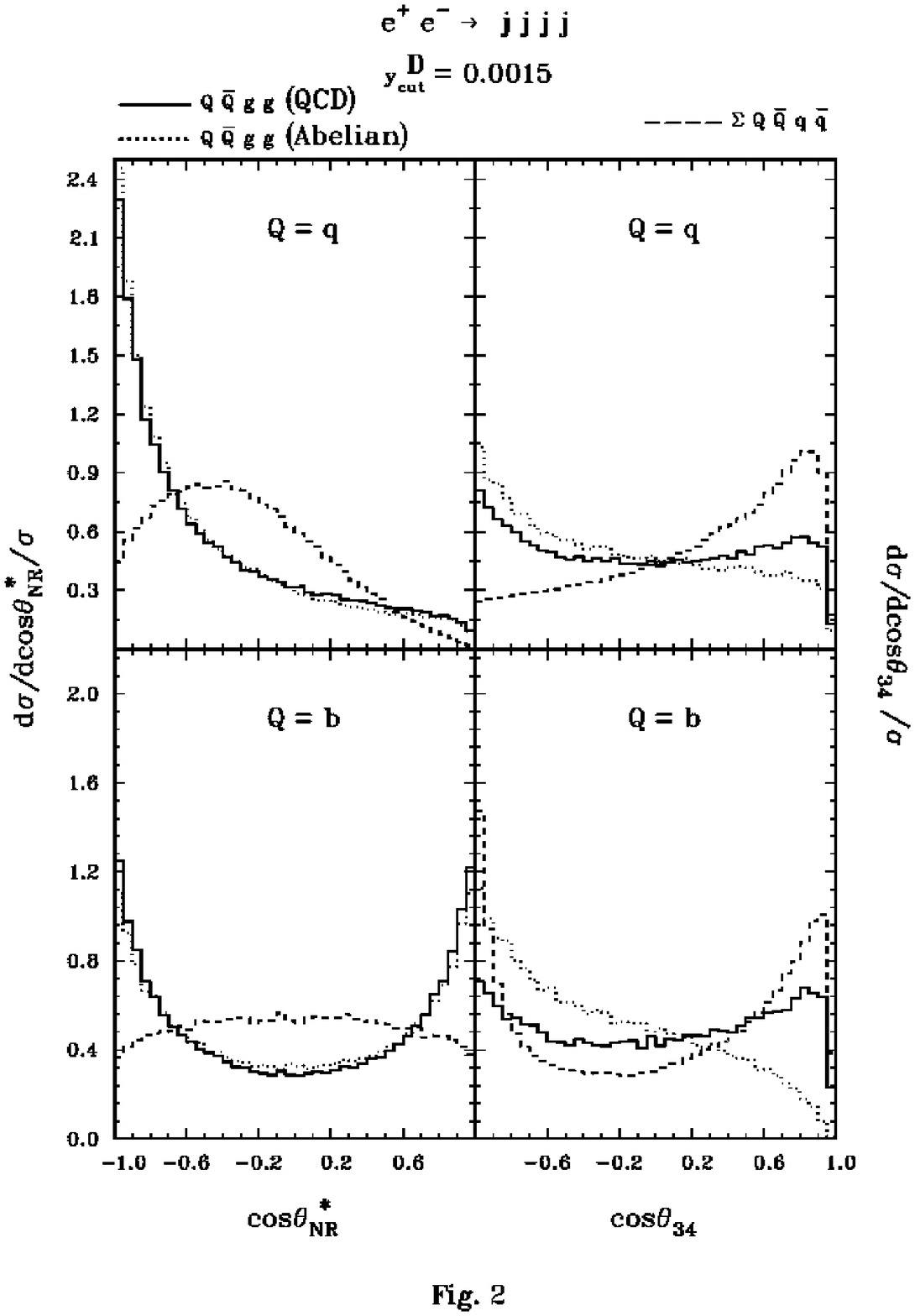,height=22cm}
\vskip0.0005cm
\noindent
{\small Distributions in the cosine of the angles 
$\theta_{NR}^*$ and $\theta_{34}$
for the $Q\bar Qgg$ (QCD and Abelian) and $Q\bar Qq\bar q$
components in 4-jet events, in the D scheme with 
$y^D_{\mathrm{cut}}=0.0015$,
without ($Q=q$)  and with ($Q=b$) $b$-tagging.}
\end{figure}
\stepcounter{figure}
\vfill
\clearpage

\begin{figure}[p]
~\epsfig{file=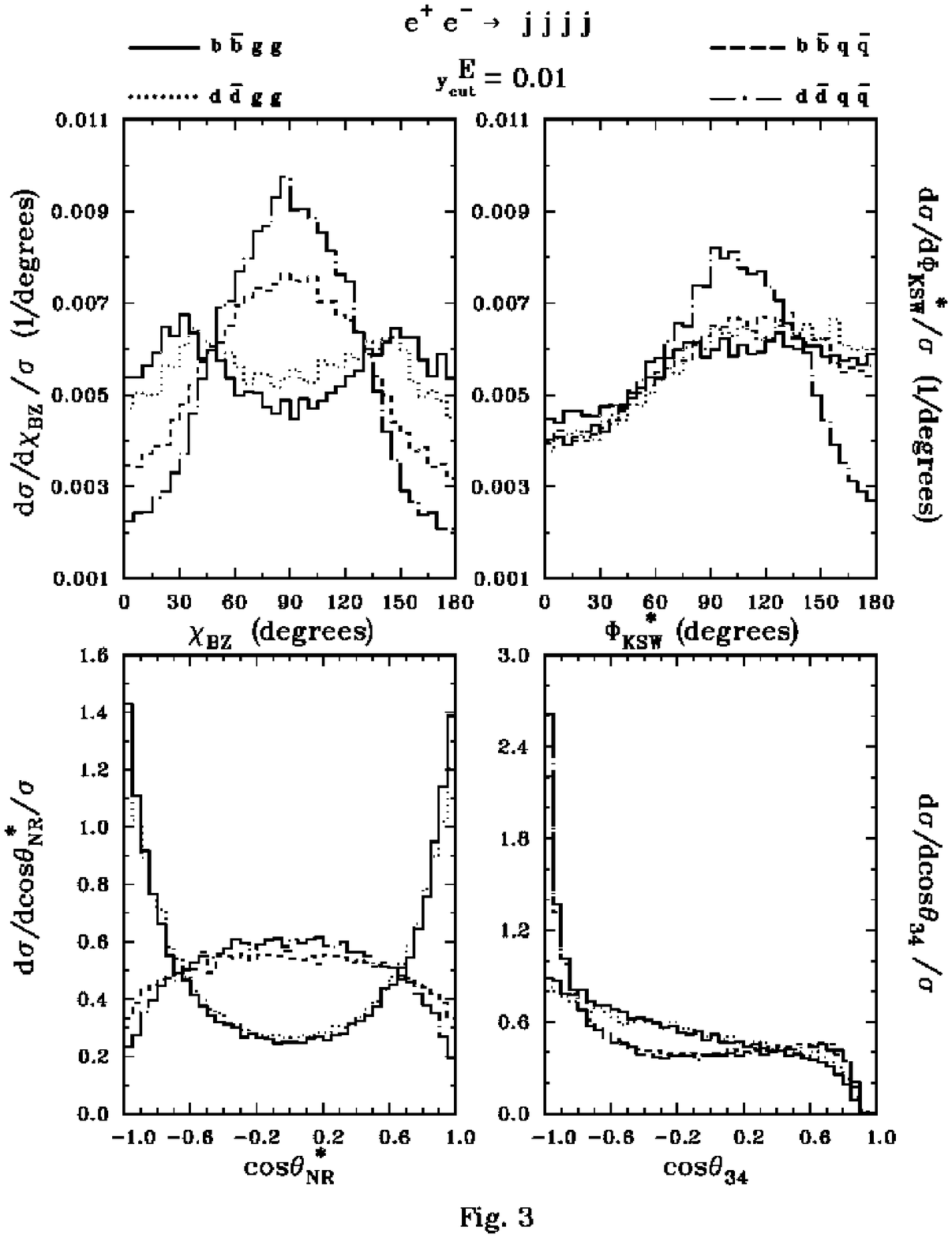,height=22cm}
\vskip0.0005cm
\noindent
{\small Distributions in the angles $\chi_{BZ}$ and $\Phi_{KSW}^*$
and in the cosine of the angles $\theta_{NR}^*$ and $\theta_{34}$,
for the $Q\bar Qgg$ 
and $Q\bar Qq\bar q$
components in 4-jet events, in the E scheme with 
$y^E_{\mathrm{cut}}=0.01$,
where $Q=b,d$ and $q$ is a massless quark.}
\end{figure}
\stepcounter{figure}
\vfill
\clearpage

\begin{figure}[p]
~\epsfig{file=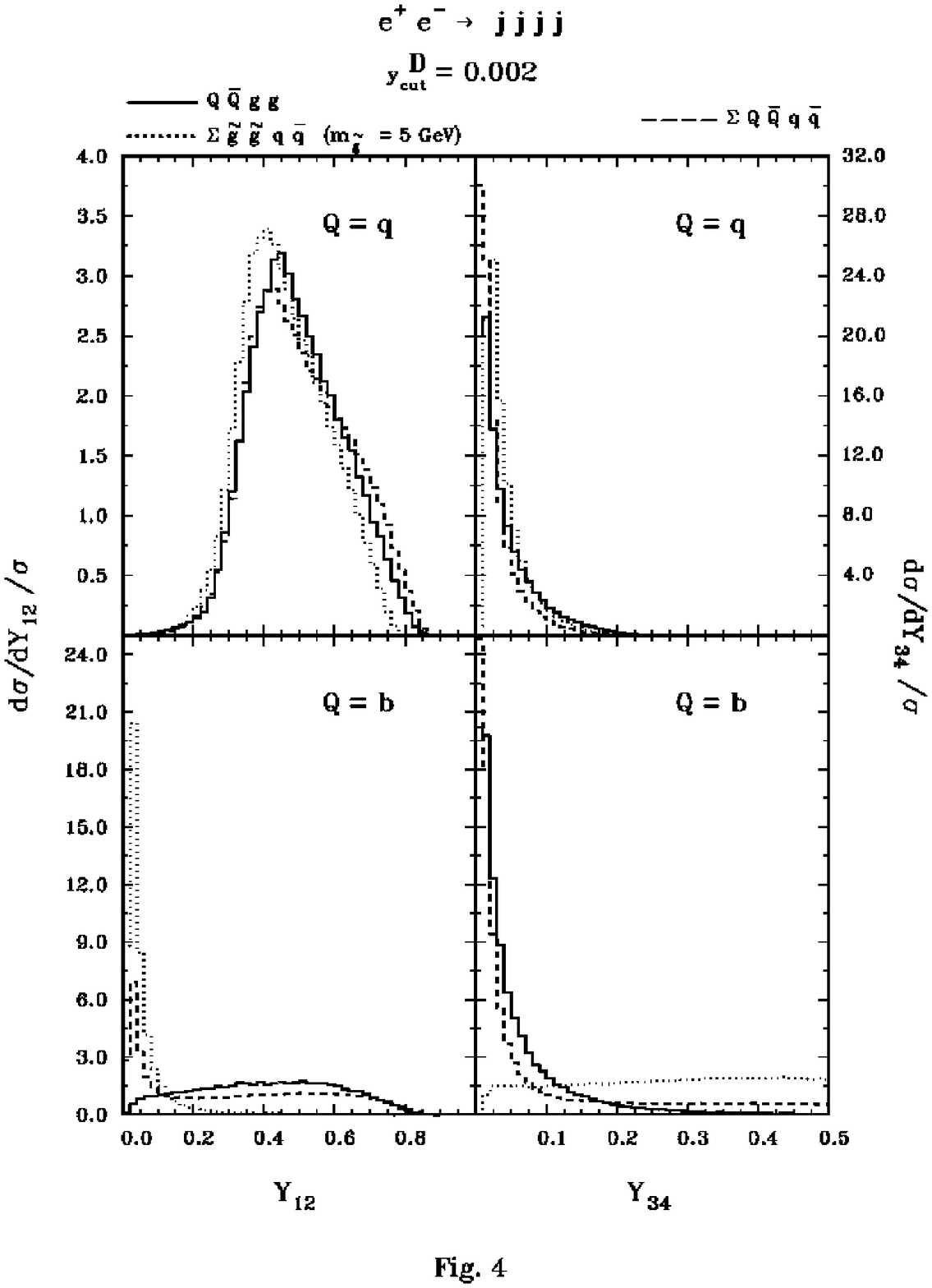,height=22cm}
\vskip0.0005cm
\noindent
{\small Distributions in the rescaled invariant masses $Y_{ij}=M^2_{ij}/s$,
where $ij=12,34$, for ordinary QCD and for SUSY 4-jet 
events, in the D scheme with 
$y^D_{\mathrm{cut}}=0.002$,
without ($Q=q$)  and with ($Q=b$) vertex tagging. Here, $\mgluino=5$ GeV.}
\end{figure}
\stepcounter{figure}
\vfill
\clearpage

\begin{figure}[p]
~\epsfig{file=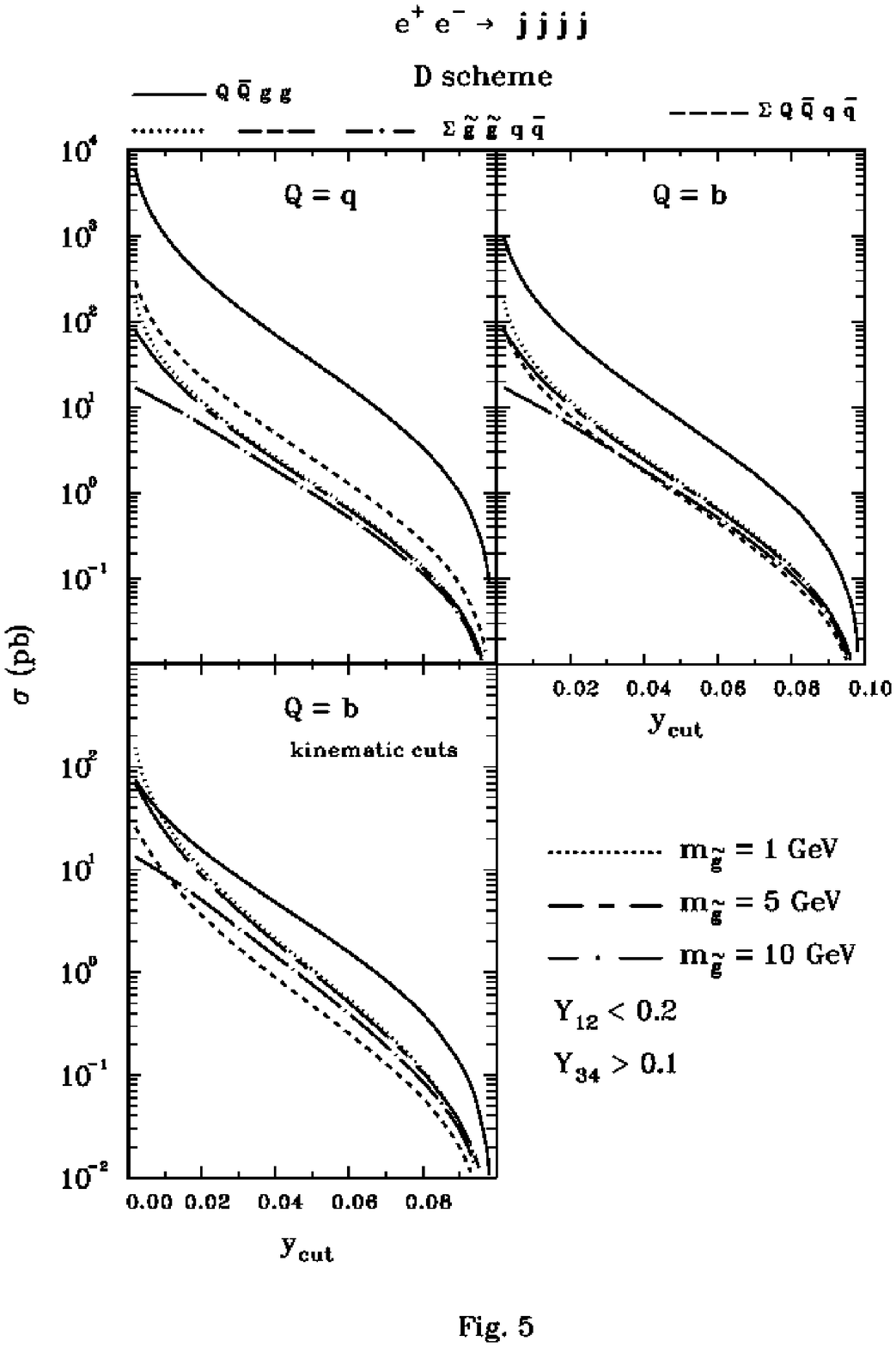,height=22cm}
\vskip0.0005cm
\noindent
{\small Total cross sections of ordinary QCD and of SUSY 4-jet 
events, in the D scheme, 
without ($Q=q$)  and with ($Q=b$) vertex tagging, 
and after the kinematic cuts, for three different values of $\mgluino$.}
\end{figure}
\stepcounter{figure}
\vfill
\clearpage

\begin{figure}[p]
~\epsfig{file=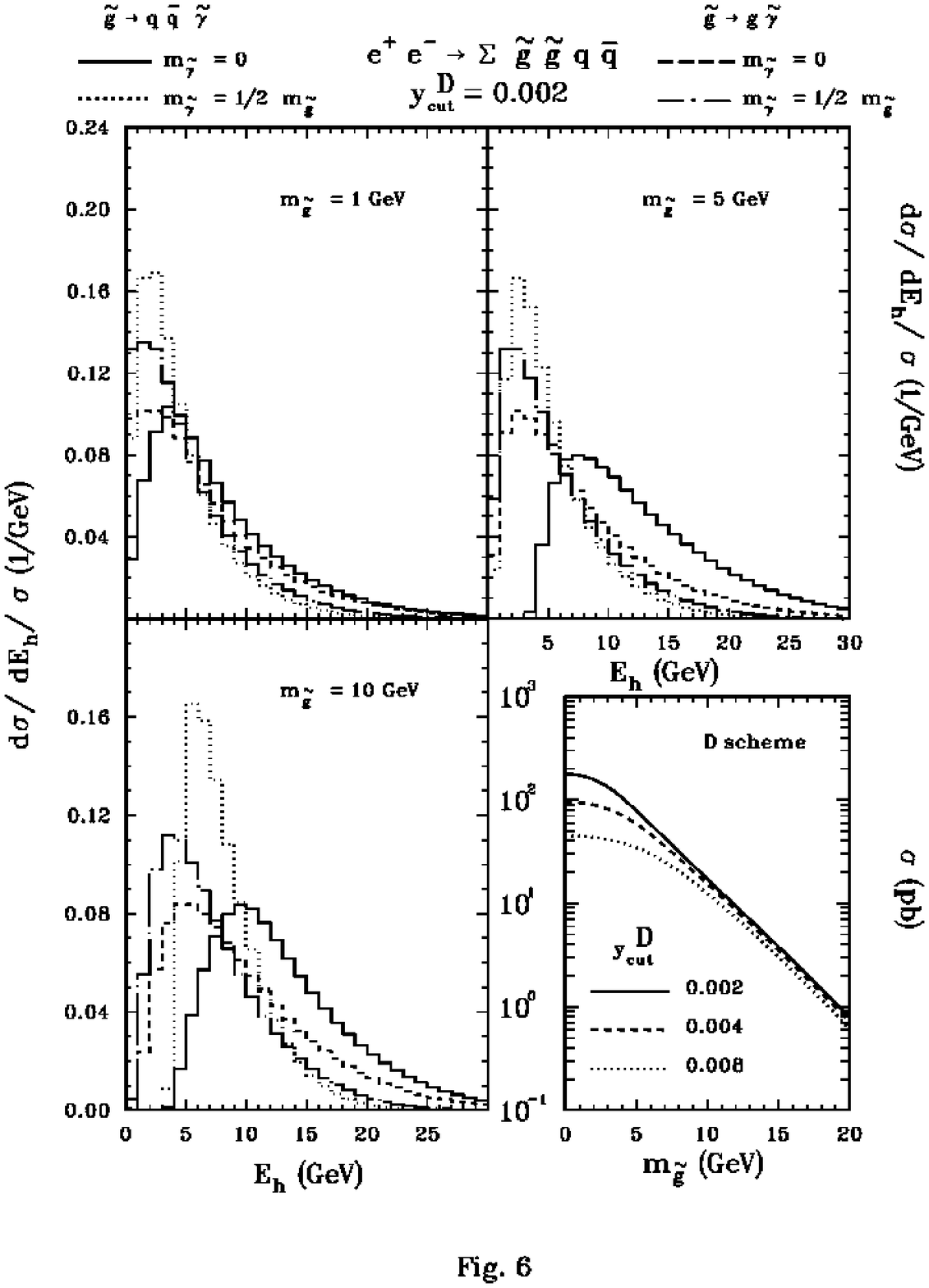,height=22cm}
\vskip0.0005cm
\noindent
{\small Differential distributions in the hadronic
energy of the `gluino jet' after the two possible SUSY decays,
in the D scheme, for various combinations of $\mgluino$ and $\mphotino$;
and total cross section of gluino events in 4-jets, as a function
of $\mgluino$ and for three different values of 
$y^D_{\mathrm{cut}}$.}
\end{figure}
\stepcounter{figure}
\vfill
\clearpage

\end{document}